\documentclass[%
 preprint,
 amsmath,amssymb,
 aps,
pra,
]{revtex4-2}
\usepackage{graphicx}% Include figure files
\usepackage{dcolumn}% Align table columns on decimal point
\usepackage{bm}% bold math

\begin{document}

\title{Optical polarization singularities in metallic cavities excited by electric dipole sources}

\author{Shiqi Jia} \affiliation{Department of Physics, City University of Hong Kong, Tat Chee Avenue, Kowloon, Hong Kong, China} 

\author{Tong Fu} \affiliation{Department of Physics, City University of Hong Kong, Tat Chee Avenue, Kowloon, Hong Kong, China} 

\author{Shubo Wang}\email {shubwang@cityu.edu.hk} \affiliation{Department of Physics, City University of Hong Kong, Tat Chee Avenue, Kowloon, Hong Kong, China} 

\date{\today}% It is always \today, today,

\begin{abstract}
Optical polarization singularities (PSs) in real space carry rich topological properties and can enable highly precise manipulations of light fields. Conventional studies focus on the PSs in the open space of optical systems. The properties of PSs inside optical cavities remain largely unexplored. By using full-wave finite-element simulations, we investigate the optical PSs inside metallic cavities excited by electric dipole sources. We determine the topological indices, morphology, and spatial evolutions of the singularities inside both spherical and torus cavities. We uncover the relationship between spatial symmetries and the PSs, as well as the mechanism underlying the emergence of polarization Möbius strips in the spherical cavity. In addition, we investigate the topological transitions of the PSs connecting two geometries (i.e., sphere and torus) with distinct topologies. The results provide insight into the singular and topological properties of light fields in optical cavities and can find applications in optical sensing and optical manipulation of small particles.
\end{abstract}

\maketitle
\newpage
%%%%%%%%%%%%%%%%%%%%%%%%%%  body  %%%%%%%%%%%%%%%%%%%%%%%%%%
\section{Introduction}
Polarization singularities (PSs) are special polarization defects that possess intricate topological properties. The PSs of optical fields were firstly discussed by Nye and Hajnal \cite{nye_wave_1987} and have attracted increasing attention recently \cite{bliokh_geometric_2019,ruchi_phase_2020,wang_polarization_2021,liu_topological_2021}. There are generally three types of polarization singularities, namely the C point, L point, and V point. The C point exhibits circular polarization with an indeterminate major axis of the polarization ellipse. The L point displays linear polarization with an indeterminate normal direction of the polarization ellipse. At the V point, the field amplitude is zero and the field direction is ill-defined. Optical PSs can be realized with various methods, including light focusing \cite{schoonover_polarization_2006}, interference \cite{larocque_reconstructing_2018}, and scattering \cite{grigoriev_fine_2018,kuznetsov_three-dimensional_2020,peng_polarization_2021}, in diverse systems such as metasurfaces \cite{zhang_generation_2019,dorrah_tunable_2022} and photonic crystals \cite{zhang_observation_2018}. The rich topological properties of the PSs give rise to intriguing physics in both momentum space and real space, such as bound states in the continuum \cite{zhen_topological_2014,doeleman_experimental_2018}, geometric phases \cite{bliokh_geometric_2019,fu_near-field_2024}, non-Hermitian exceptional points \cite{zhou_observation_2018}, skyrmions \cite{shen2021supertoroidal,gutierrez-cuevas_optical_2021,sugic_particle-like_2021,shen2024optical}, and superoscillations \cite{fu_topological_2024}. Recently, the concept of PSs is extended to acoustic waves and water waves \cite{bliokh_polarization_2021,muelas-hurtado_observation_2022,tong2024sculpturingsoundfieldsrealspace}. For three-dimensional (3D) electromagnetic fields, the PSs can form continuous lines designated as C lines, L lines, and V lines accordingly. The PS lines have been theoretically studied in various systems including paraxial waves \cite{dennis_polarization_2002,berry_electric_2004}, Gaussian beam \cite{freund_polarization_2002}, and general 3D fields \cite{berry_index_2004}. These PS lines exhibit nontrivial topological configurations in real space, including links and knots \cite{leach_knotted_2004,pisanty_knotting_2019,larocque_optical_2020}. It was shown that the polarization ellipses, when traced along a closed path encircling a C line, can form a Möbius strip with intriguing topological properties \cite{garcia-etxarri_optical_2017,kuznetsov_topology_2021,peng_topological_2022}. Since C points carry optical spin angular momentum, they can be used for chiral detection and chiral discrimination \cite{jia_chiral_2022,jia_broadband_2023}.

In 3D space, a general monochromatic magnetic field can be expressed as $\mathbf{H}=(\mathbf{A}+$ $i \mathbf{B}) \exp (i \theta)$, where $\mathbf{A}$ and $\mathbf{B}$ are the major axis and the minor axis of the polarization ellipse, respectively \cite{dennis_polarization_2002,berry_index_2004}. The $\theta$ corresponds to the phase of the scalar field $\Psi=\mathbf{H} \cdot \mathbf{H}$ with $\theta=\operatorname{arg}(\Psi) / 2$. The phase $\theta$ is ill-defined at the C points, i.e., the C points correspond to the phase singularities of the scalar field $\Psi$. Therefore, the topological properties of C points can be characterized by a quantized phase index $I_{\mathrm{ph}}=1 /(2 \pi) \oint \nabla \arg (\Psi) \cdot d \mathbf{r}$. It is notable that the sign of $I_{\mathrm{ph}}$ depends on the direction of the integration loop. In addition, a polarization index $I_{\mathrm{pl}}=1 /(4 \pi) \oint d \varphi$ with $\varphi$ being the azimuthal angle on the Poincaré sphere, which corresponds to the winding number of the major axis $\mathbf{A}$ of the polarization ellipse, can be assigned to each C point. For the C points exhibiting a texture of star, monstar, and lemon types, their polarization indices are $-1 / 2,1 / 2$, and $1 / 2$, respectively \cite{nye_wave_1987}. Notably, the phase and polarization indices are related by $I_{\mathrm{pl}}=\operatorname{sign}(\mathbf{t} \cdot \mathbf{S}) I_{\mathrm{ph}} / 2$, where $\mathbf{t}$ is a tangent unit vector of the C line and $\mathbf{S}=\operatorname{Im}\left(\mathbf{H}^* \times \mathbf{H}\right)$ is the local magnetic spin \cite{peng_topological_2022}. These properties and definitions also apply to general electric fields in 3D space.

In the far field of a scattering system, the electric and magnetic fields are tangent to the spherical wavefront and the momentum sphere. The corresponding polarization fields can be considered tangent line fields defined on a smooth manifold (i.e., spherical surface). In accordance with the Poincaré-Hopf (PH) theorem, the summation of the topological indices of the PSs in the tangent fields is solely determined by the topology (i.e., Euler characteristic) of the manifold, thereby ensuring the existence of PSs in the far field \cite{chen_singularities_2019,peng_polarization_2021}. Consequently, the global topological properties (i.e., total index) of the PSs are the same in the far field, despite the variations in the scatterers’ geometry and incident light properties. On the other hand, the properties of the PSs in the near field heavily depend on the geometry and symmetry of optical systems, which have attracted significant attention recently and led to intriguing physics associated with the real-space topology, with promising applications in encoding information \cite{larocque_optical_2020}, optical sensing \cite{jia_chiral_2022,jia_broadband_2023}, and optical metrology \cite{yuan_detecting_2019}. These studies have focused on the PSs in the open space with radiation boundary condition, where the PSs' properties are mainly attributed to the scattering fields. The PSs in closed spaces such as optical cavities remain largely unexplored. Investigation of these PSs can uncover intriguing topological properties of the optical cavity modes.

In this paper, we investigate the properties of magnetic PSs in metallic cavities induced by electric dipole sources. We focus on magnetic PSs because the magnetic field is tangent to the cavity boundary and its polarization represents a tangent line field. We employ full wave simulations to determine the magnetic PSs in spherical and torus cavities with Euler characteristics $\chi=2$ and $\chi=0$, respectively. Akin to the near-field PSs  in open space, the sum of the polarization indices of all the PSs on the cavity inner surface is bounded and equal to the Euler characteristic of the cavity, as dictated by the PH theorem. We find that the detailed configuration of the PSs strongly depends on the excitation and system symmetry. We observe that the mirror symmetry can protect the emergence of V points, while the cylindrical symmetry consistently leads to PS lines with polarization index $I_{\mathrm{pl}}=+1$, including the V lines and higher-order C lines. The impact of symmetry on the polarization Möbius strips of various orders in the spherical cavity is further discussed. In addition, we uncover the nontrivial topological transition of PS lines connecting the surfaces of two distinct topological manifolds (i.e., sphere and torus). 

The paper is organized as follows. In Sec. II, we present and discuss the results of the magnetic PSs in a spherical cavity excited by electric dipole sources and the PSs in the eigen fields of the spherical cavity. For comparison, in Sec. III, we discuss the magnetic PSs in a torus cavity excited by electric dipole sources, which possess different global topological properties due to the different Euler characteristic of torus. In Sec. IV, we show the polarization Möbius strips in the spherical cavity and uncover the topological transition of PS lines connecting the outer surface of a metallic torus and the inner surface of a spherical cavity, which helps to illustrate the synergy between different real-space topology of the cavities. We draw the conclusion in Sec. V.

\section{Polarization singularities in a spherical cavity}

\begin{figure}[htb]
\centering\includegraphics{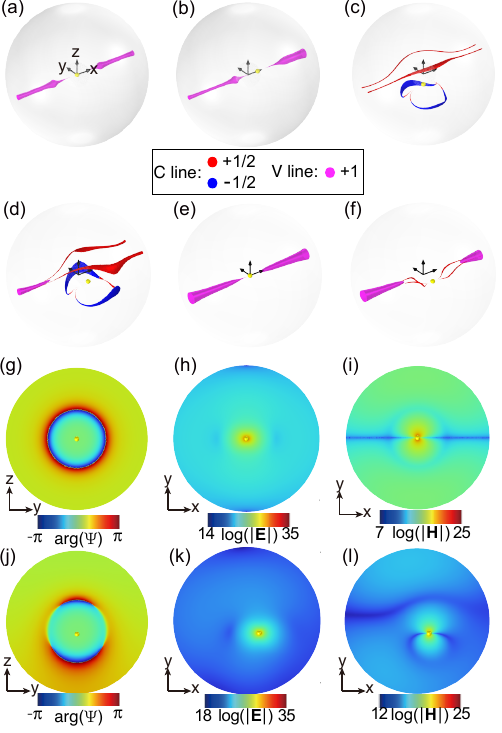}
\caption{The PSs in the spherical cavity excited by a linearly polarized electric dipole $\mathbf{p}=p \hat{\mathbf{x}}$. (a-d) C lines (blue/red) and V lines (magenta) generated by the dipole source locating at (0, 0, 0), $(50~ \mu \mathrm{m}, 0, 0)$, $(0, 0, -50~ \mu \mathrm{m})$ and $(50~ \mu \mathrm{m}, 0, -50~ \mu \mathrm{m})$, respectively. The frequency is $f = 500$ GHz. (e, f) C lines (red) and V lines (green) at the lowest resonance frequency $f = 436$ GHz. The locations of the dipole are the same as in (a) and (d). (g-i) The $\arg (\Psi)$, electric field amplitude, and magnetic field amplitude in the case of (a). (j-l) The $\arg (\Psi)$, electric field amplitude, and magnetic field amplitude in the case of (d).}
\label{fig1}
\end{figure}

We consider a copper spherical cavity with a radius of $300~ \mu \mathrm{m}$ excited by an internally positioned linear electric dipole $\mathbf{p}=p \hat{\mathbf{x}}$ at the frequency of $f=500$ 
GHz. The boundary condition imposed by copper ensures that the magnetic field is nearly parallel to the surface of the cavity. We conduct full wave numerical simulation of the system by using a finite element package COMSOL. Figure \ref{fig1}(a) shows the PS lines (magenta colored) emerging inside the cavity, where the electric dipole source (denoted by the yellow dot) locates at the origin. Due to the cylindrical symmetry with respect to the $x$ axis and the mirror symmetry with respect to the $yoz$ plane, the PS lines exhibit a simple configuration of straight lines, which coincide with the $x$ axis and connect the opposite surfaces of the cavity and the dipole source. The PS lines are V lines with polarization index $I_{\mathrm{pl}}=+1$. Therefore, the total polarization index of the two V points on the inner spherical surface is $\sum I_{\mathrm{pl}}=+2$, which equals the Euler characteristic of sphere (i.e., $\chi=2$ ). This result is not accidental and can be understood as follows. Since the magnetic field is parallel to the cavity inner surface, its polarization distribution on the surface can be considered a line field defined on a smooth manifold (i.e., the spherical surface). According to the PH theorem, such a line field must have discrete singularities carrying a net topological index equal to the Euler characteristic of the manifold \cite{needham_visual_2021}. These singularities are just the polarization singularities on the inner surface of the spherical cavity.

To understand the dependence of the PSs on the excitation and system symmetry, we determine the PS lines induced by the same electric dipole source locating at different positions inside the cavity, as illustrated in Figs. \ref{fig1}(b) to 1(d). In Fig. \ref{fig1}(b), the dipole source locates at $(50~ \mu \mathrm{m}, 0,0)$, which breaks the mirror symmetry respect to the $yoz$ plane but maintains the cylindrical symmetry with respect to the $x$ axis. We observe that the two V lines of index $I_{\mathrm{pl}}=+1$ remain aligned with the $x$ axis but are asymmetric with respect to the $yoz$ plane. This indicates that the cylindrical symmetry protects the V lines. In Fig. \ref{fig1}(c), the dipole source locates at $(0,0,-50~ \mu \mathrm{m})$, breaking the cylindrical symmetry. As expected, the V lines disappear, and two C lines with index $I_{\mathrm{pl}}=+\frac{1}{2}$ (red) and two C lines with index $I_{\mathrm{pl}}=-\frac{1}{2}$ (blue) appear symmetrically inside the cavity instead. In this case, there are four C points on the cavity surface. The sum of their indices is $\sum I_{\mathrm{pl}}=4 \times \frac{1}{2}=+2$, which is the same as the cases of Figs. \ref{fig1}(a) and \ref{fig1}(b) and consistent with the PH theorem. Figure \ref{fig1}(d) shows the PS lines for the dipole source locating at $(50~ \mu \mathrm{m},0,-50~ \mu \mathrm{m})$, which breaks the cylindrical symmetry and the mirror symmetry with respect to $yoz$ plane. Different from the case of Fig. \ref{fig1}(c), there are two C lines merging into a V line with index $I_{\mathrm{pl}}=+1$. Thus, there are one V point with $I_{\mathrm{pl}}=+1$ and two C points with $I_{\mathrm{pl}}=+1/2$ on the cavity surface, leading to the same total index $\sum I_{\mathrm{pl}}=2\times1/2+1=+2$. These results demonstrate that the configuration and index of individual PS line strongly depend on the excitation and system symmetry, but their global topological property (i.e., total index of all surface singularities) is only decided by the Euler characteristic of the cavity.

The spherical cavity supports multiple resonance modes at the excitation frequency, and their interferences give rise to the PS lines in Figs. \ref{fig1}(a)-\ref{fig1}(d). The complexity of the PS lines is highly related to the number of excited modes and their relative amplitudes. A dominating resonance mode at low frequencies usually leads to simple configurations of PS lines. An example is shown in Fig. \ref{fig1}(e), where the electric dipole locates at the origin and excites the lowest resonance mode (i.e., linear electric dipole mode)   of the cavity. We can see that the PS lines (i.e., V lines) are similar to the case of Fig. \ref{fig1}(a). Notably, when the electric dipole source locates at $(50~ \mu \mathrm{m},0,-50~ \mu \mathrm{m})$, as shown in Fig. \ref{fig1}(f), only parts of the V lines bifurcate into C lines, in contrast to the case of Fig. \ref{fig1}(d). The higher stability of the V lines is attributed to the linear polarization of the cavity resonance mode.

To understand the field properties associated with the PSs (i.e., V points and C points) inside the cavity, we evaluate the phase $\arg (\Psi)=\arg (\mathbf{H} \cdot \mathbf{H})$ on the $zoy$ plane for the cases in Figs. \ref{fig1} (a) and \ref{fig1}(d). The results are shown in Figs. \ref{fig1}(g) and \ref{fig1}(j), respectively. We noticed that no phase singularity appears in Fig. \ref{fig1}(g), while two pairs of phase singularities with phase index $I_{\mathrm{ph}}= \pm 1$ appear in Fig. \ref{fig1}(j) at the positions of the C points. This is because V points do not necessarily exhibit singular phase. In contrast, at the C points, $\Psi=\mathbf{H} \cdot \mathbf{H}=\left(|\mathbf{A}|^2-|\mathbf{B}|^2\right) \mathrm{e}^{i 2 \theta}=0$ as a result of $|\mathbf{A}|=|\mathbf{B}|$, and $\arg (\Psi)$ is always ill-defined. In Figs. \ref{fig1}(h) and \ref{fig1}(k) we show the electric field amplitude on the $xoz$ plane for the cases in Figs. \ref{fig1}(a) and \ref{fig1}(d), respectively. Fig.s \ref{fig1}(i) and \ref{fig1}(l) show the corresponding magnetic field amplitude. It can be clearly seen that, at the position of the V lines, the magnetic field vanishes while the electric field remains finite.

\begin{figure}[htb]
\centering\includegraphics{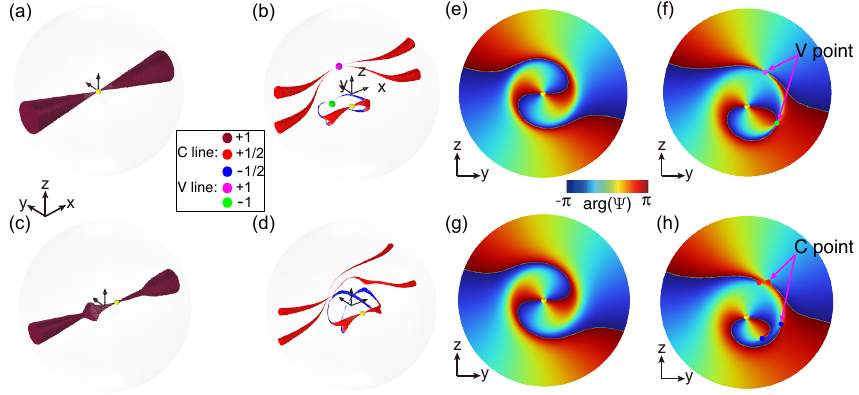}
\caption{The PSs in the spherical cavity induced by a circularly polarized electric dipole $\mathbf{p}=p(\hat{\mathbf{y}}+i \hat{\mathbf{z}})$. (a-d) C lines (blue/red/dark-red) and V lines (magenta/green) induced by the dipole source at (0,0,0), $(0,0,-50~ \mu \mathrm{m})$, $(50~ \mu \mathrm{m},0,0)$, and $(50~ \mu \mathrm{m}$, 0, $-50~ \mu \mathrm{m}$), respectively. (e-h) The distribution of $\arg (\Psi)$ corresponding to the cases of (a-d).}
\label{fig2}
\end{figure}

While the global topological property of the surface PSs does not depend on specific excitation, the dipole source can affect the types and configurations of the PSs. As a demonstration, we consider the PSs excited by a circularly polarized electric dipole $\mathbf{p}=p(\hat{\mathbf{y}}+i \hat{\mathbf{z}})$. Figure \ref{fig2}(a) shows the PSs when the circular dipole locates at the origin. The system has the cylindrical symmetry with respect to $x$ axis and the mirror symmetry with respect to the $y o z$ plane. Due to these symmetries, two higher order C lines with $I_{\mathrm{pl}}=+1$ (colored in dark red) emerge, connecting the dipole and the cavity surface. The sum of the indices for the two C points on the cavity surface is still +2 , thereby satisfying the PH theorem. Figure \ref{fig2}(b) shows the PSs when the circular electric dipole source locates at $(0,0,-50~ \mu \mathrm{m})$, which breaks the cylindrical symmetry. In this case, only the lowest order C lines with $I_{\mathrm{pl}}= \pm 1 / 2$ appear. In addition, there are two V points carrying opposite polarization index (green and magenta colored) appear due to the mirror symmetry with respect to the $y o z$ plane. The V point with $I_{\mathrm{pl}}=$ +1 (denoted by the magenta point) corresponds to the crossing of two C lines with $I_{\mathrm{pl}}=+1 / 2$, while the V point with $I_{\mathrm{pl}}=-1$ (denoted by the green point) corresponds to the crossing of two C lines with $I_{\mathrm{pl}}=-1 / 2$. Figure \ref{fig2}(c) shows the case with the source at ( $50~ \mu \mathrm{m}, 0,0$ ), which maintains the cylindrical symmetry but breaks the mirror symmetry with respect to the $yoz$ plane. Different from the case in Fig. \ref{fig2}(a), the higher order C lines become asymmetrically distributed. Finally, we set the dipole source at ($50~ \mu \mathrm{m}$, $0,-50~ \mu \mathrm{m})$ so that both the cylindrical and mirror symmetries are absent. In this case, multiple C lines with complex configurations emerge in the cavity, as shown in Fig. \ref{fig2}(d). Notably, there are only four C points $I_{\mathrm{pl}}=+1 / 2$ on the cavity surface, and their total index satisfies the PH theorem.

Figs. \ref{fig2}(e)-\ref{fig2}(h) show the phase arg $(\Psi)$ on the $yz$ cross sections at the location of the circular dipole source for the cases of Figs. \ref{fig2}(a)-\ref{fig2}(d). The vortex patterns of the phase distribution in Figs. \ref{fig2}(e) and \ref{fig2}(g) arise from the cylindrical symmetry of the systems. It is noteworthy that the V points exhibit a phase index $I_{\mathrm{ph}}= \pm 2$, which is consistent with the polarization index $I_{\mathrm{pl}}$ through the relation $I_{\mathrm{pl}}=\operatorname{sign}(\mathbf{t} \cdot \mathbf{S}) I_{\mathrm{ph}} / 2$. Figures \ref{fig2}(f) and \ref{fig2}(h) show the phase $\arg (\Psi)$ for the cases in Figs. \ref{fig2}(b) and \ref{fig2}(d), respectively. Due to the breaking of the mirror symmetry, the V point with $I_{\mathrm{ph}}=+2\left(I_{\mathrm{ph}}=-2\right)$ in Fig. \ref{fig2}(f), marked as magenta (green) dot, splits into two C points with $I_{\mathrm{ph}}=+1\left(I_{\mathrm{ph}}=-1\right)$, marked as red (blue) dots in Fig. \ref{fig2}(h). Clearly, the phase index is conserved after the splitting. This demonstrates that the V points are equivalent to two C points that are spatially degenerate, which are generally unstable under perturbations.

\begin{figure}[htb]
\centering\includegraphics{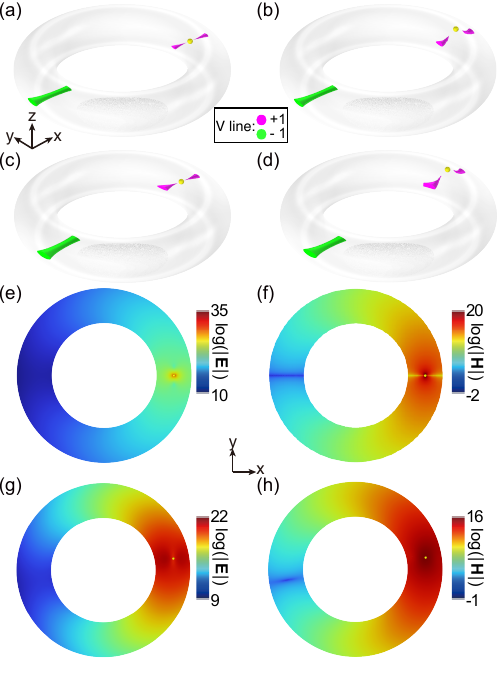}
\caption{The PSs in the torus cavity induced by the electric dipole $\mathbf{p}=p \hat{\mathbf{x}}$ at $f=500$ GHz. (a-d) The V lines induced by the dipole at $(600~ \mu \mathrm{m}, 0,0)$,$(600~ \mu \mathrm{m}, 0, 100~ \mu \mathrm{m})$, $(600~ \mu \mathrm{m}, 100~ \mu \mathrm{m}, 0)$ and $(600~ \mu \mathrm{m}, 100~ \mu \mathrm{m}, 100~ \mu \mathrm{m})$. (e, f) Electric and magnetic field amplitudes on the $xoy$ plane with the dipole at $(600~ \mu \mathrm{m}, 0,0)$. (g, h) Electric and magnetic field amplitudes with the dipole at $(600~ \mu \mathrm{m}, 100~ \mu \mathrm{m}, 100~ \mu \mathrm{m})$.}
\label{fig3}
\end{figure}

\section{Polarization singularities in a torus cavity}
According to the PH theorem, the global topological property of the tangent line field is solely decided by the Euler characteristic of the manifold. Thus, it is essential to explore the optical polarization singularities in the cavities with different Euler characteristics. We consider a copper torus cavity with a major radius of $600~ \mu \mathrm{m}$ and a minor radius of $150~ \mu \mathrm{m}$, which has the Euler characteristic $\chi=0$. Figure \ref{fig3} shows the PSs induced by the electric dipole $\mathbf{p}=p \hat{\mathbf{x}}$ (denoted by the small yellow dot) in the torus cavity at $f=500$ GHz. Figure \ref{fig3}(a) shows the PSs for the dipole positioned at ($600~ \mu \mathrm{m}$, 0, 0 ). We notice that two V lines with $I_{\mathrm{pl}}=+1$ and $I_{\mathrm{pl}}=-1$ appear symmetrically along the $x$ direction and connect to the inner surface of the cavity. This can be attributed to the mirror symmetries of the system with respect to the $xoz$ and $yoz$ planes. Importantly, the sum of the polarization indices of the four V points on the cavity surface is zero, agreeing with the Euler characteristic $\chi=0$ of torus and satisfying the PH theorem. We further simulate the PSs excited by the dipole source locating at different positions $(600~ \mu \mathrm{m}, 0,100~ \mu \mathrm{m}),(600~ \mu \mathrm{m}, 100~ \mu \mathrm{m}, 0)$, and $(600~ \mu \mathrm{m}, 100~ \mu \mathrm{m}, 100~ \mu \mathrm{m})$. The results are illustrated in Figs. \ref{fig3}(b), \ref{fig3}(c), and \ref{fig3}(d), respectively. As seen, the PS lines are slightly distorted due to the breaking of symmetries, but their main features remain unchanged, i.e. the two V lines carrying opposite polarization indices appear in the opposing parts of the torus cavity. The simple and stable configurations of the V lines can be attributed to the subwavelength size of the torus cross section. At the working frequency $f=500$ GHz, which is below the cutoff frequency of the fundamental waveguide mode supported by the torus cavity, the field inside the cavity is evanescent field with a quasi-static property, leading to the simple polarization configuration.

Figures \ref{fig3}(e) and \ref{fig3}(f) show the amplitudes of the electric and magnetic fields on the $xoy$ plane inside the torus cavity, corresponding to the case of Fig. \ref{fig3}(a). It can be observed that the magnetic V lines correspond to local minima of the magnetic field amplitude, while the electric field amplitude is not affected by the V lines. Figures \ref{fig3}(g) and \ref{fig3}(h) show the amplitudes of the electric and magnetic fields on the $xoy$ plane for the case in Fig. \ref{fig3}(d). As seen, the magnetic field on the right part of the torus cavity is strongly influenced by the source location, i.e., it does not show a vanished amplitude due to the distortion of the V lines. In contrast, the magnetic field pattern in the left part is only slightly shifted, and the dark line with near zero amplitude can still be observed clearly.

\begin{figure}[htb]
\centering\includegraphics{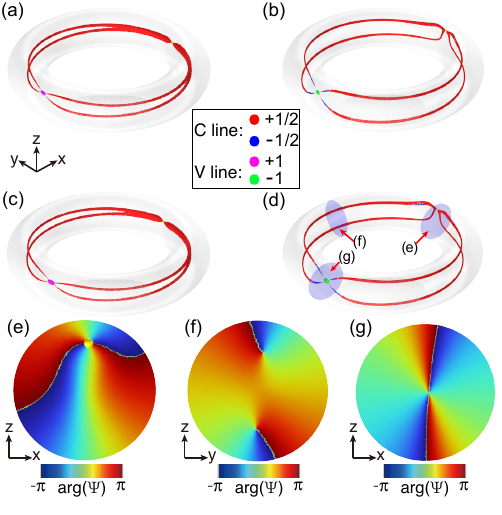}
\caption{The PSs in the torus cavity induced by the circularly polarized electric dipole $\mathbf{p}=p(\hat{\mathbf{x}}+i \hat{\mathbf{z}})$ at $f=500$ GHz. (a-d) The C lines induced by the dipole at $(600~ \mu \mathrm{m}, 0,0),(600~ \mu \mathrm{m}, 0,100~ \mu \mathrm{m})$, $(600~ \mu \mathrm{m}, 100~ \mu \mathrm{m}, 0)$, and $(600~ \mu \mathrm{m}, 100~ \mu \mathrm{~m}, 100~ \mu \mathrm{m})$, respectively. (e-g) The distribution of $\arg (\Psi)$ on the cross sections marked in (d).}
\label{fig4}
\end{figure}

We also simulate the PSs in the torus cavity excited by a circularly polarized electric dipole $\mathbf{p}=p(\hat{\mathbf{x}}+i \hat{\mathbf{z}})$ at the frequency $f=500$ GHz. The results are shown in Figs. \ref{fig4}(a)-\ref{fig4}(d) for the dipole source locating at the positions $(600~ \mu \mathrm{m}, 0,0),(600~ \mu \mathrm{m}, 0,100~ \mu \mathrm{m})$, $(600~ \mu \mathrm{m}, 100~ \mu \mathrm{m}, 0)$, and $(600~ \mu \mathrm{m}, 100~ \mu \mathrm{~m}, 100~ \mu \mathrm{m})$, respectively. Naturally, the circular electric dipole can generate two higher order C lines with $I_{\mathrm{pl}}=+1$ along the $\hat{\mathbf{y}}$ direction. However, due to the absence of cylindrical symmetry in the torus cavity, the high-order C lines bifurcate into lowest order C lines with $I_{\mathrm{pl}}=+1 / 2$. Interestingly, these C lines do not connect to the torus surface but form rings inside the cavity. Thus, the total polarization index on the cavity surface is zero for all the cases in Figs. \ref{fig4}(a) to \ref{fig4}(d), agreeing with the PH theorem. Since the system in Fig.  \ref{fig4}(a) has a mirror symmetry with respect to the $x o z$ plane, the C lines intersect and form a V point with $I_{\mathrm{pl}}=+1$ (denoted by the magenta dot). For the cases with breaking symmetries, corresponding to Figs.  \ref{fig4}(b)-\ref{fig4}(d), the C lines also form C rings and intersect to give a V point. Notably, the V points in Figs.  \ref{fig4}(b) and \ref{fig4}(d) carry a negative index $I_{\mathrm{pl}}=-1$, different from the cases in Figs.  \ref{fig4}(a) and \ref{fig4}(c). This is attributed to a sign flip of the polarization index of the two C lines, as shown by the transition from red color to blue color.

To further understand the phase properties of the magnetic field in the cavity, we plot $\arg (\Psi)$ on the three cross sections labeled as "(e)", "(f)", and "(g)" in Fig.  \ref{fig4}(d). The phase distributions are depicted in Figs.  \ref{fig4}(e), (f), and (g), respectively. It is evident that a higher order phase singularity with the phase index $I_{\mathrm{ph}}=+2$ appears precisely at the location of the dipole source. For the case in Fig.  \ref{fig4}(f), the two phase singularities correspond to the two C lines with polarization index $I_{\mathrm{pl}}=+1 / 2$ in Fig.  \ref{fig4}(d). For the case in Fig. $4(\mathrm{g})$, we notice a phase singularity with $I_{\mathrm{ph}}=-2$, corresponding to the V point in Fig.  \ref{fig4}(d).

\section{Möbius strip and topological transition}

\begin{figure}[htb]
\centering\includegraphics{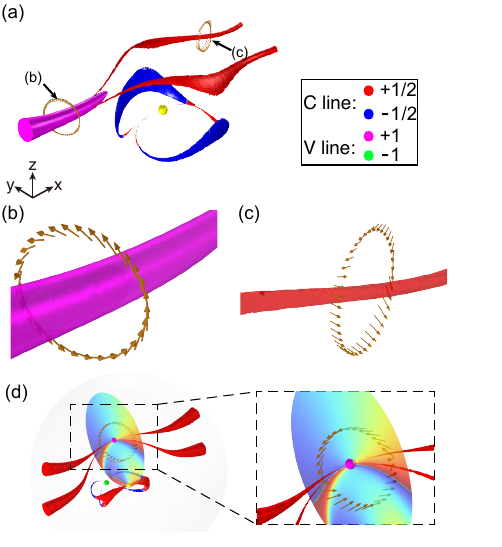}
\caption{(a) The Möbius strip for the loops enclosing different PS lines in the spherical cavity. (b) No Möbius strip appears for the loop enclosing the single V line. (c) One twist Möbius strip for the loop enclosing a single C line. (d) A double-twist Möbius strip for the loop enclosing a mirror-symmetry-protected V point.}
\label{fig5}
\end{figure}

In addition to the phase and amplitude of the internal fields, the PSs strongly affect the polarization distribution and can give rise to interesting polarization structures. It is widely acknowledged that polarization Möbius strips can emerge around a single C line in free space \cite{garcia-etxarri_optical_2017}. Generally, a polarization strip with an odd number of twists exhibits inherent topological complexity, while a strip with an even number of twists can always be transformed into an untwisted strip featuring trivially aligned polarizations \cite{bliokh_geometric_2019}. Here, we explore the polarization Möbius strips on the loops enclosing different PS lines within the spherical cavity, as shown in Fig. \ref{fig5}(a), which corresponds to the case in Fig. \ref{fig1}(d). We consider one loop enclosing the V line with $I_{\mathrm{pl}}=+1$ and the C line with $I_{\mathrm{pl}}=+1 / 2$. Figs. \ref{fig5}(b) and \ref{fig5}(c) show the zoom-ins of the two polarization strips. As seen, the polarization strip of the V lines is untwisted and does not form a nontrivial Möbius strip. This is because that the V line here is equivalent to two degenerate C lines with opposite spin, and the phase index of the V line is $I_{\mathrm{ph}}=0$. In contrast, the loop enclosing the C line exhibits a Möbius strip with one-turn twisting. While the V line in Fig. \ref{fig5}(b) cannot induce nontrivial polarization Möbius strips, the mirror symmetry can give rise to double-twist Möbius strips \cite{peng_topological_2022}. An example is the V point in Fig. \ref{fig2}(b). For a loop enclosing the V point and normally piercing the mirror plane, the polarization major axis $\mathbf{A}$ forms a mirror-symmetric double-twist Möbius strip, as shown in Fig. \ref{fig5}(d). The inset of Fig. \ref{fig5}(d) shows that $\mathbf{A}$ undergoes a single twist on each side of the mirror plane, thereby forming a symmetrical double-twist Möbius strip. The phase $\arg (\Psi)$ is shown on the mirror plane. If the mirror symmetry is broken, the C lines will be gapped, and this double-twist Möbius strip can transform into a trivial strip with no twist, corresponding to the case in Fig. \ref{fig2}(d).

\begin{figure}[htb]
\centering\includegraphics{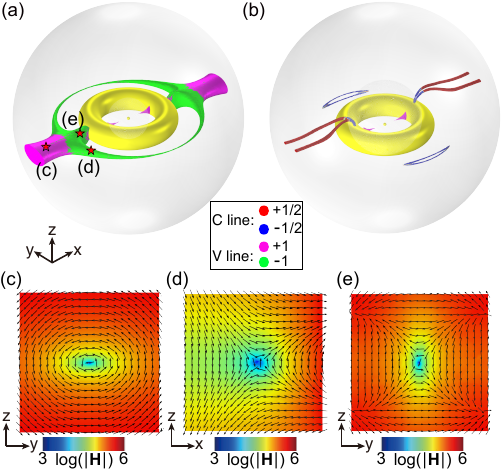}
\caption{(a) Topological transition of the PS lines in the sphere-torus system with mirror symmetries. (b) Topological transition of the PS lines in the sphere-torus system without the mirror symmetry with respect to the $xoy$ plane. (c-e) Magnetic field amplitude and polarization distribution at three different cut planes marked by the red stars in (a).}
\label{fig6}
\end{figure}

The PS lines can exhibit interesting topological transitions due to the interaction of the cavities with different topology. We consider a spherical cavity with a radius of 300 $\mu \mathrm{m}$, which encloses a torus cavity with a major radius of $100~ \mu \mathrm{m}$ and a minor radius of $30~ \mu \mathrm{m}$, as depicted in Fig. \ref{fig6}(a). The torus and sphere share the same geometric center. Due to the different Euler characteristics of the sphere and torus, the total indices of the PSs are different on the inner surface of the sphere and on the outer surface of the torus. This indicates that a topological transition must happen to the PS lines in the space sandwiched by the two geometries. Figure \ref{fig6}(a) shows the PS lines excited by the electric dipole source $\mathbf{p}=p \hat{\mathbf{x}}$ positioned at the center at $f=500$ GHz. We notice that only V lines emerge in the cavity. In particular, the V lines connecting the torus and the sphere display mirror symmetry and undergo an index change from -1 to +1 as they extend from the outer surface of the torus to the inner surface of the spherical cavity. Such a transition is necessary to guarantee the PSs on both surfaces satisfy the PH theorem and can be understood as follows. The V lines generated by electric dipole source in the center region of the torus are mainly determined by the electric dipole source itself, which carry the index $I_{\mathrm{pl}}=+1$. To satisfy the PH theorem on torus surface, an equal number of V lines with $I_{\mathrm{pl}}=-1$ must emerge at outer surface of torus. Meanwhile, the PSs on the inner surface of spherical cavity must also satisfy the PH theorem, i.e., their total polarization index must be +2. Therefore, it is necessary for the V lines to undergo a transition of the topological index in the middle region. To understand the field properties near the transition point, we plot in Figs. \ref{fig6}(c)-\ref{fig6}(e) the magnetic field amplitude and polarization major axis $\mathbf{A}$ on three cutting planes marked by the red stars in Fig. \ref{fig6}(a). We see that the magnetic field vanishes at the V points and the polarization winding number agrees with the index of the V lines.

We note that different types of topological transitions can happen in the sphere-torus system if different excitations are employed. Figure \ref{fig6}(b) shows the case where the $xoy$ mirror symmetry is broken by changing the location of torus and the dipole source to $(0,0,20~ \mu \mathrm{m})$ in the cavity. As seen in Fig. \ref{fig6}(b), the V lines near the electric dipole have the same configuration as in Fig. \ref{fig6}(a). However, the PS lines connecting the two surfaces are four C lines undergoing an index change from -1/2 to +1/2, which is different from the case in Fig. \ref{fig6}(a). In this case, the unstable V lines transform into C lines due to the breaking of the mirror symmetry. Importantly, the PSs on both the spherical and torus surfaces still satisfy the PH theorem.

\section{Conclusion}
We investigate the characteristics of magnetic PSs in metallic cavities induced by electric dipole sources, including topological indices, morphology, and spatial evolutions. By considering two topologically distinct cavities (i.e., spherical and torus cavities), we uncover the impact of structural topology on the properties of the PSs. We find that the difference in the cavity topology will dramatically affect the configuration of PS lines, but the sum of the polarization topological indices of all surface PSs is always equal to the Euler characteristic of the cavity, as governed by the PH theorem. By tailoring the position and polarization of the electric dipole sources, we control the symmetry of the systems and uncover the relationship between the spatial symmetries and the PSs. We found that the mirror and cylindrical symmetries can induce higher order PSs, like the V lines and higher order C lines. We further discuss the properties of the polarization Möbius strips emerging in the spherical cavity and the topological transition between two cavities with different topology. The study contributes to the understanding of the interactions between structured light fields and optical cavities. The results can find applications in chiral light-matter interactions such as cavity-enhanced chiral sensing and discriminations. The structured light fields in the optical cavities may also be utilized for optical manipulations of small particles. 

\section*{Acknowledgments} 
The work described in this paper was supported by grants from the National Natural Science Foundation of China (No. 12322416) and the Research Grants Council of the Hong Kong Special Administrative Region, China (No. AoE/P-502/20).

%%%%%%%%%%%%%%%%%%%%%%% References %%%%%%%%%%%%%%%%%%%%%%%%%

%%%%%%%%%% If using BibTeX:
\bibliography{sample}

%%%%%%%%%% If preparing manually:
% \newcommand{\enquote}[1]{``#1''}

% \bibitem{Zhang:14}
% Y.~Zhang, S.~Qiao, L.~Sun, Q.~W. Shi, W.~Huang, L.~Li, and Z.~Yang,
%   \enquote{Photoinduced active terahertz metamaterials with nanostructured
%   vanadium dioxide film deposited by sol-gel method,}
%   {\protect\JournalTitle{Optics Express}} \textbf{22}, 11070--11078 (2014).

% {Optical Society}, \enquote{{OSA Publishing},}
%   \url{http://www.osapublishing.org}.

% \bibitem{FORSTER2007}
% P.~Forster, V.~Ramaswamy, P.~Artaxo, T.~Bernsten, R.~Betts, D.~Fahey,
%   J.~Haywood, J.~Lean, D.~Lowe, G.~Myhre, J.~Nganga, R.~Prinn, G.~Raga,
%   M.~Schulz, and R.~V. Dorland, \enquote{Changes in atmospheric consituents and
%   in radiative forcing,} in \enquote{Climate Change 2007: The Physical Science
%   Basis. Contribution of Working Group 1 to the Fourth assesment report of
%   Intergovernmental Panel on Climate Change,}  S.~Solomon, D.~Qin, M.~Manning,
%   Z.~Chen, M.~Marquis, K.~B. Averyt, M.~Tignor, and H.~L. Miler, eds.
%   (Cambridge University Press, 2007).

% \end{thebibliography}

\end{document}